\documentclass[%
 reprint,
 amsmath,amssymb,
 aps,
prl,
]{revtex4-2}

\usepackage{graphicx}
\usepackage{dcolumn}
\usepackage{bm}
\usepackage[dvipsnames]{xcolor}
\usepackage[normalem]{ulem}
\usepackage{float}

\begin{document}


\title{Landau-Levich Enhanced Cheerios Effect}
\author{Hadrien Bense}\email{hadrien.bense@ulb.be}
\author{Emmanuel Siéfert}\email{emmanuel.siefert@ulb.be}
 \author{Fabian Brau}\email{fabian.brau@ulb.be}
\affiliation{Université libre de Bruxelles (ULB), Nonlinear Physical Chemistry Unit, CP231, 1050 Bruxelles, Belgium}%

\date{\today}
\begin{abstract}
We study the capillary attraction force between two fibers dynamically withdrawn from a bath. We propose an experimental method to measure this force and show that its magnitude strongly increases with the retraction speed by up to a factor ten compared to the static case. We show that this remarkable increase stems from the shape of the dynamical meniscus between the two fibers. We first study the dynamical meniscus around one fiber, and obtain experimental and numerical scaling of its size increase with the capillary number, which is not captured by the classical Landau-Levich-Derjaguin theory. We then show that the shape of the deformed air-liquid interface around two fibers can be inferred from the linear superposition of the interface around a single fiber. These results yield an analytical expression for the attraction which compares well with the experimental data. Our study reveals the critical role of the retraction speed to create stronger capillary interactions, with potential applications in industry or biology.
\end{abstract}

\maketitle

\begin{figure}[t]
\centering
\includegraphics[width=\columnwidth]{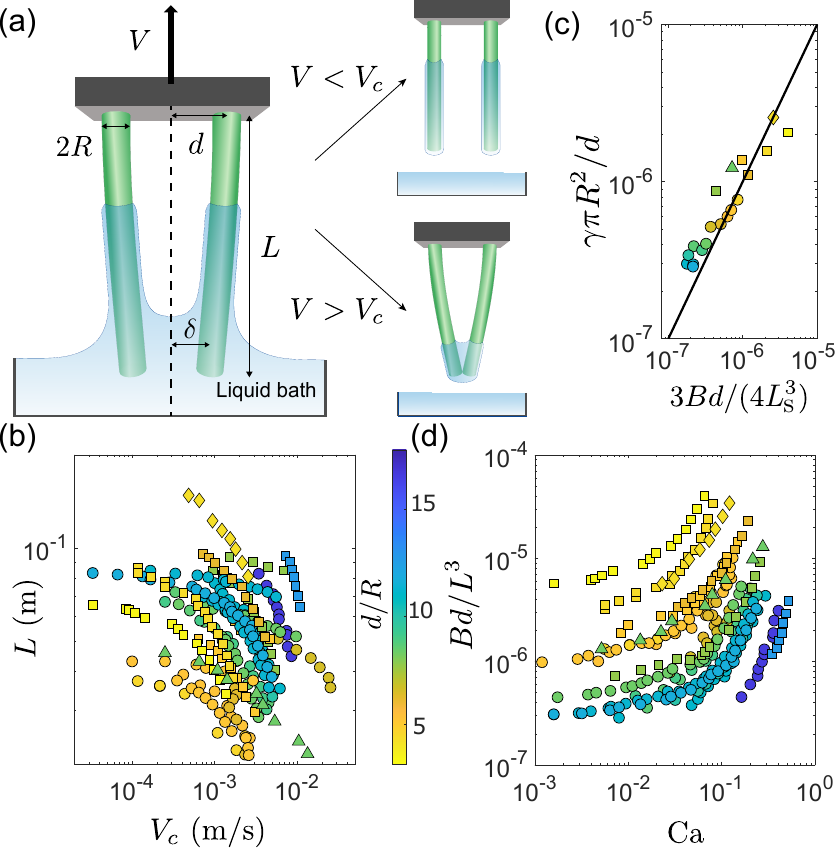}
\caption{(a) Experimental set up. Two parallel glass fibers ($E= 64\pm1$ GPa, radius $R$, length $L$), separated by a distance $2d$ and clamped at one end, are removed from a fluid bath at a speed $V$. Above a critical velocity $V_c$, the fibers coalesce. (b) Length $L$ of the fibers as a function of the critical velocity $V_c$. Squares: $R=100$ $\mu$m; circles: $R=50$ $\mu$m; diamonds: $R=160$ $\mu$m, all in silicon oil ($\mu = 0.96$ or $0.096$ Pa s). Triangles: $R=50$ $\mu$m in glycerol. Color: $d/R$ given by the colorbar. (c) At $V=0$ m/s, the scaling for $L=L_{\text{s}}$ obtained in Ref.~\cite{siefert2022capillary} is recovered (solid curve). (d) Estimation of the dynamic capillary force $Bd/L^3$ as a function of $\text{Ca}$.
}
\label{fig1}
\end{figure}

Dip coating, a technique consisting in drawing an object out of a liquid bath~\cite{ruschak1985coating,ruschak2004coating,schweizer2012liquid}, is used in industry to functionalize objects with e.g., antireflective~\cite{jonsson2010effect}, biocompatible~\cite{mohseni2014comparative}, hydrophobic~\cite{kapridaki2013tio2} or electrothermal~\cite{janas2014review} coatings. This process is also used by some animals, such as bees, bats or birds, to feed by dipping their brush-like tongue in nectar~\cite{lechantre2021essential,Harper2013,Mitchell1990}. Each object, when dipped in a liquid, deforms the interface which induces a force acting on its neighbors. This long-range capillary interaction, known as the ``Cheerios effect''~\cite{nicolson1949interaction,kralchevsky1992capillary,kralchevsky2001capillary,vella2005cheerios,ho2019direct,botto2012capillary,mcgorty2010colloidal,cooray2012capillary}, is mediated by the shape of the meniscus surrounding the objects. Beyond its fundamental interest, the Cheerios effect has been recently harnessed in the context of particle self-assembly~\cite{botto2012capillary,mcgorty2010colloidal,zeng20223d}. However, the effect of dynamics on the Cheerios effect, when, e.g. dip coated structures are swiftly removed from a bath, remains poorly investigated. Dip coating is usually studied with the Landau-Levich-Derjaguin (LLD) theory, which predicts that for thin fibers removed at sufficiently small speed from a bath of Newtonian liquid, this thickness is given by~\cite{derjaguin1943thickness,deryagin1963theory,quere1999fluid,zhang2022dip}:
\begin{equation}
\label{t-LLD}
   t=1.34\, R\, \text{Ca}^{2/3},
\end{equation}
where $R$ is the fiber radius, $\text{Ca}=\mu V/\gamma$ the capillary number, $\mu$ the viscosity, $\gamma$ the surface tension of the liquid, $V$ the withdrawal speed. This relation has been obtained in a regime where the Reynolds number, $\text{Re} = \rho V R/\mu$, and the Bond number, $\text{Bo} = R^2/\ell_c^2$, are small ($\ell_c=\sqrt{\gamma/\rho_{\ell} g}$ is the capillary length, $\rho_{\ell}$ the liquid density and $g$ the gravitational acceleration). This relation was proved successful and has been extended to larger $\text{Ca}$, $\text{Bo}$ and $\text{Re}$~\cite{Tallmadge1965,white1965,white1966theory,Burkina1980,Wilson1982,quere95,quere96,quere98inertia,Benilov2008} and to different liquids such as partially-wetting~\cite{snoeijer2008thick}, suspension~\cite{gans2019dip,palma2019dip}, non-Newtonian~\cite{tanguy1984finite,afanasiev2007landau,ashmore2008coating}, polymer~\cite{quere98,zhang22b} and surfactant~\cite{quere97,shen2002fiber,mayer2012} solutions (see also the recent review~\cite{rio2017withdrawing}). The LLD model essentially assumes that the thin lubricated film deposited on the fiber connects to a static meniscus. One may thus naively think that pulling two fibers out of a bath should barely modify the air-liquid interface compared to its static shape, and thus the force between two objects. However, experiments show that the attraction force between two fibers increases by up to a factor ten with $\text{Ca}$ (Fig.~\ref{fig1}).

In this Letter, we rationalize the capillary menisci-mediated interaction between two fibers as they are dynamically withdrawn from a liquid bath. We develop an original measurement technique, based on the deflection of elastic fibers, to experimentally access the dynamical capillary force. We show that it may be one order of magnitude larger than the static one computed in Ref.~\cite{kralchevsky1992capillary}. To understand this large increase of the force, we analyze the dynamical meniscus around a single withdrawn fiber. We show that its significant deviation from the static one can be accounted by rescaling  its height far enough from the fiber by a factor depending on $\text{Ca}$. We also show that, in good approximation, the dynamical meniscus around two fibers is the sum of the dynamical menisci around the same isolated fibers. Therefore, the dynamical meniscus around two fibers can be reconstructed from the static meniscus around a single fiber. We finally draw analogy from the static case, to show that the force directly stems from distortion of the air-liquid interface and propose a scaling law accounting for the large force observed. 

\begin{figure}[t]
\centering
\includegraphics[width=\columnwidth]{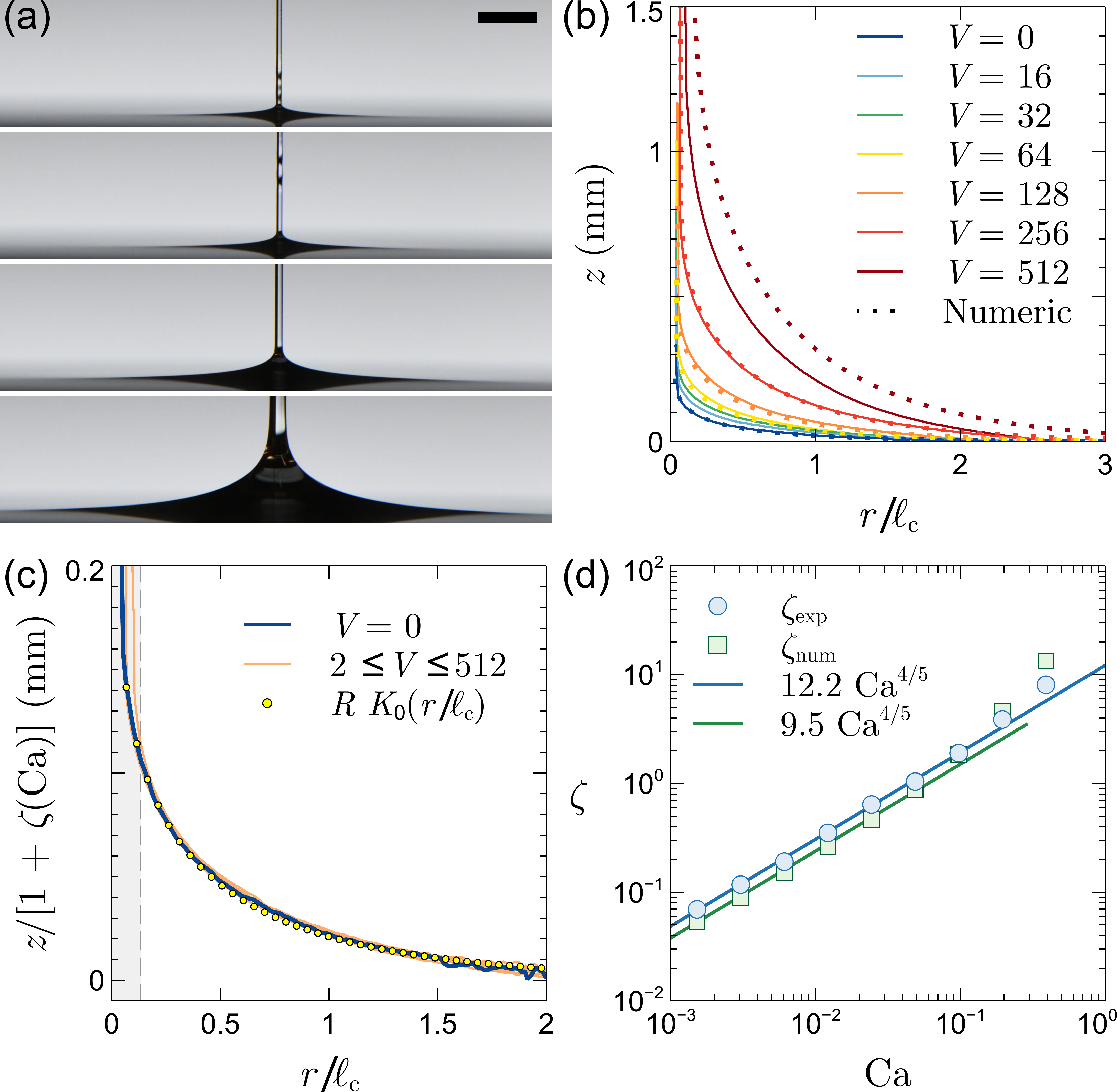}
\caption{(a) Snapshots of a glass fiber ($R=50$ $\mu$m) withdrawn from silicon oil ($\mu=0.96$ Pa s) at various retraction speeds ($V=0$, $16$, $128$, $512$ mm/min). Scale bar: 1 mm. (b) Meniscus shape as a function of the rescaled radial distance $r/\ell_c$ at various $V$ (mm/min). The agreement with the theoretical model proposed in Ref.~\cite{zhang2022dip} is good up to $V=256$ mm/min ($\text{Ca}\simeq 0.2$). (c) Collapse of the dynamical meniscus profiles onto the static meniscus shape when their height are properly rescaled by $1+\zeta$. $V=2^n$ mm/min with $1 \le n \le 9$. (d) Evolution of the scaling factor $\zeta$ as a function of the capillary number. $\zeta_{\text{exp}}$ and $\zeta_{\text{num}}$ are the scaling factors for the experimental and numerical profiles respectively~\cite{sup}.}
\label{fig2}
\end{figure}

To perform the experiments, two identical fibers of length $L$ and radius $R$ are clamped vertically to a linear stage at a distance $2d$ and immersed in a silicone oil bath~\cite{sup}. The fibers are first quasi-statically removed from the bath to find the dry length $L_{\text{0}}$ at which they coalesce~\cite{siefert2022capillary}. The fibers are then cut at a length $L<L_{\text{0}}$ to prevent them from coalescing at $\text{Ca}=0$. The fibers are then re-immersed and dynamically withdrawn from the bath at various increasing retraction speeds until they coalesce at $V=V_c$ (Fig.~\ref{fig1}(a)). This process of shortening the fibers and finding $V_c$ is repeated until the fibers are too short to coalesce at any retraction speed considered ($\text{Re} \lesssim 10^{-2}$, $\text{Ca} \lesssim 0.5$). Shortening the fibers allows us to ensure that, when coalescence occurs, the immersed part of both fibers is sufficiently small to neglect any viscous drag acting on the structures. This iterative process yields the critical speed $V_c$ at which the two fibers coalesce as a function of $L$. Figure~\ref{fig1}(b) shows that $L$ decreases as $V$ increases so that the Cheerios effect between the two fibers is enhanced by their dynamical withdrawal. 

As shown in Ref.~\cite{siefert2022capillary} for the quasi-static case, coalescence occurs when the capillary force, $F_{\text{s}}=\gamma\pi R^2/\delta$, equals the elastic force, $F_{\text{el}} = 3B \delta/L^3$, required to bend the fibers over a distance $\delta=d/2$. As seen in Fig.~\ref{fig1}(c), the scaling $F_{\text{s}}(d/2)=F_{\text{el}}(d/2)$ is recovered at $\text{Ca}=0$. This scaling suggests to plot $B{d}/{L^3}$ as a function of the capillary number $\text{Ca}$ to estimate the dynamic capillary force $F$. Figure~\ref{fig1}(d) shows that the latter decreases as $d/R$ increases like in the static case but increases up to one order of magnitude when the capillary number increases at fixed $d/R$.

The interaction between the fibers is mediated by the meniscus created around them~\cite{siefert2022capillary,kralchevsky1992capillary,cooray2012capillary}. To get some insight into the deformation of the air-liquid interface as the fibers are withdrawn from the bath, the dynamical meniscus around a single fiber is first analyzed. The latter grows significantly with the withdrawal speed as shown in Fig.~\ref{fig2}(a),(b). However, Fig.~\ref{fig2}(c) shows that rescaling the menisci height by $1+\zeta$ leads to a collapse of the profiles onto the static meniscus which, in turn, is well approximated by the modified Bessel function of the second kind of zeroth order~\cite{nicolson1949interaction,cooray2012capillary,NIST,Lo1983}. The collapse is satisfactory provided $r\gtrsim 4R$ for the investigated retraction speeds. A similar collapse is obtained for the theoretical profiles computed with the model proposed in Ref.~\cite{zhang2022dip} (see~\cite{sup}). The evolution of the scaling factor $\zeta$ as a function of $\text{Ca}$ is shown in Fig.~\ref{fig2}(d) for the experimental and numerical profiles. They both follow the same scaling law, $\zeta \sim \text{Ca}^{4/5}$, yet with a slightly different pre-factor. The size increase of the meniscus is related to the film thickening~\cite{sup}. Indeed, the static meniscus needs to connect to the fiber whose effective radius is increased by the deposited film. Since the meniscus height at a given distance far enough from the fiber is proportional to the fiber radius, i.e. $z \simeq R K_0(r/\ell_c)$, it grows with the film thickness $t$. For the values of $\text{Ca}$ considered here, $t$ increases faster than $\text{Ca}^{2/3}$, leading to a larger exponent for $\zeta$, i.e. $4/5$. As shown below, the increase of $t$ near the fiber is thus responsible for the growth of the attractive force between the fibers at much larger distances.

\begin{figure}[t]
\centering
\includegraphics[width=\columnwidth]{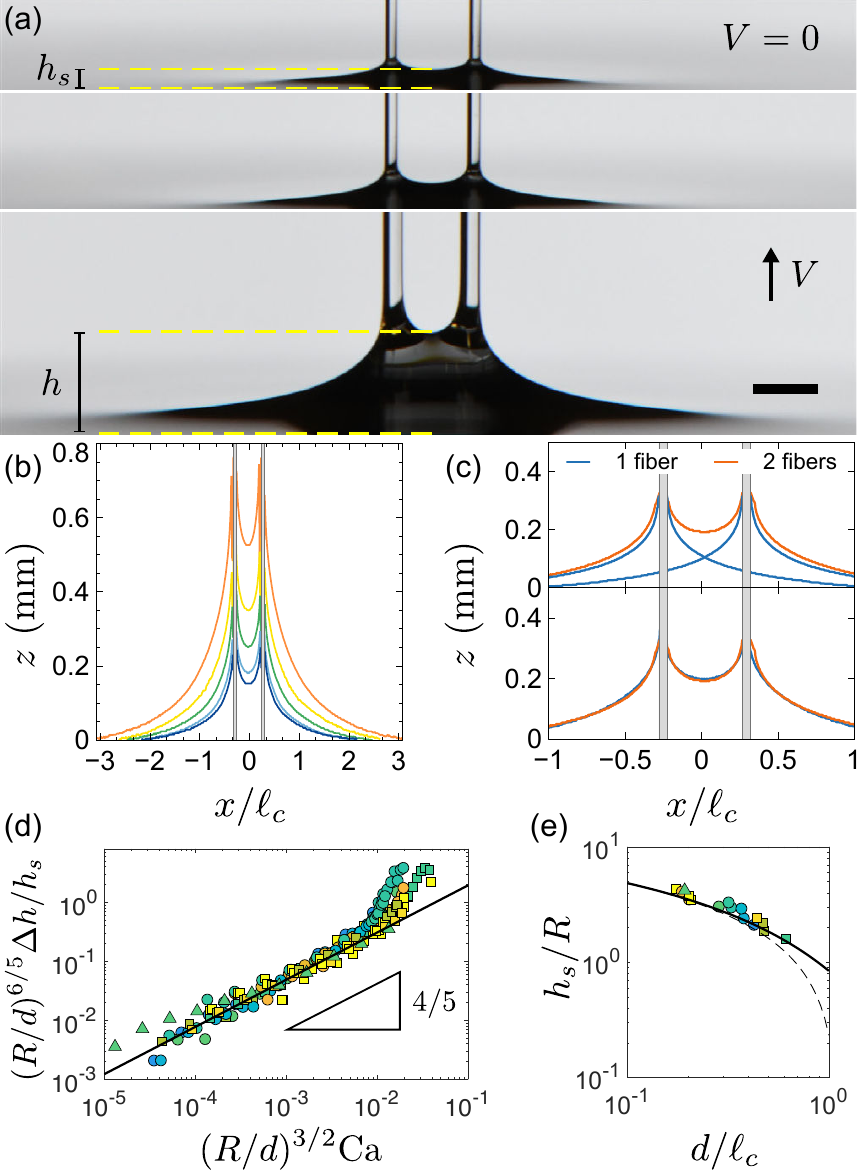}
\caption{(a) Snapshots of two glass fibers ($R=100$ $\mu$m) withdrawn from silicon oil ($\mu=0.96$ Pa s) at $V=0$, $32$, $256$ mm/min. Scale bar: 1 mm. (b) Meniscus profiles as $V$ increases ($0$, $8$, $32$, $64$, $128$ mm/min). (c) Meniscus around two fibers ($R=50$ $\mu$m) separated by a distance $2d = 0.88$ mm and withdrawn at $V=16$ mm/min ($\text{Ca} =0.012$) reconstructed from the meniscus around a single fiber withdrawn at the same speed. (d) Evolution of $\Delta h/h_{\text{s}}$ as a function of $\text{Ca}$. All data collapse on the power law obtained in Fig.~\ref{fig2}(d): $\Delta h/h_{\text{s}}=12.2\, \text{Ca}^{4/5}$. An increase of slope is observed when $d\simeq 20 R\, \text{Ca}^{2/3}$. (e) Height of the static bridge as a function of the fibers distance. The solid curve correspond to Eq.~(\ref{eq-h0}) and the dashed curve to its logarithmic approximation valid for $d/\ell_c \ll 1$. See Fig.~\ref{fig1} for the symbol and color code.}
\label{fig3}
\end{figure}

At this stage, a fairly good description of the dynamical meniscus around a single fiber has been obtained from the corresponding static one. Two identical clamped fibers are used to analyze the dynamical meniscus between them. They are sufficiently short so that no significant deflection occurs during the withdrawal and the measurements are then performed in a stationary regime (Fig.~\ref{fig3}(a),(b)). For the range of inter-fiber distances and retraction speeds considered here, the profile of the dynamical meniscus around two fibers, $z_2$, is given in good approximation by $z_2(x,y)=z_1(x-d,y) + z_1(x+d,y)$ where $z_1$ is the profile of the dynamical meniscus around a single fiber centered at $x=y=0$~ (see Fig.~\ref{fig3}(c) and \cite{sup}). This observation is strengthened by the systematic measurement of the height $h=z_2(0,0)$ of the liquid bridge for various $R$, $d$, $\gamma$ and $\mu$. According to the results reported in Fig.~\ref{fig2}, if a linear superposition of the dynamical meniscus around a single fiber applies, we should have
\begin{subequations}
\label{eq-h-all}
\begin{align}
\label{eq-h}
h &=h_{\text{s}}\, (1+\zeta), \quad \zeta \simeq 12.2\, \text{Ca}^{4/5}, \\
\label{eq-h0}
h_{\text{s}} &=2R\, K_0(d/\ell_c) \simeq 2R\, \ln[2\, \ell_c/(\gamma_{e} d)],
\end{align}
\end{subequations}
where $h_{\text{s}}$ is the height at $\text{Ca}=0$ and where the second expression in Eq.~(\ref{eq-h0}) is obtained from the expansion of $K_0$ for $d\ll \ell_c$ with $\gamma_e \simeq 1.781$ the exponential of the Euler-Mascheroni constant. Figure~\ref{fig3}(d) shows indeed that the relative variation of the bridge height, $\Delta h/h_{\text{s}}$, varies in good approximation as $\zeta$. Note that the data in Fig.~\ref{fig3}(d) has been further rescaled by powers of $R/d$. Indeed, for a given system, there is a critical retraction speed beyond which a liquid column is entrained between both fibers leading to a larger increase of $h$. This phenomenon does not appear at a given value of $\text{Ca}$ but instead when the coating film thickness on each fibers reaches a fraction of the distance $d$, i.e. when $(R/d)^{3/2} \text{Ca} \simeq 10^{-2}$, and induces a noticeable change of slope, see Fig.~\ref{fig3}(d). This explains the necessity to use an additional rescaling to capture this change of slope and obtain a good collapse of the data. Note also that Eq.~(\ref{eq-h0}) compares well with the data displayed in Fig.~\ref{fig3}(e).

At this stage, a fairly good description of the dynamical meniscus around two fibers has been obtained from the static meniscus around a single fiber. Thanks to this relationship, the dynamical force acting on two fibers during their withdrawal, and reported in Fig.~\ref{fig1}(d), can be rationalized in a simple way. For that purpose, we follow the approach developed in Ref.~\cite{kralchevsky1992capillary} to compute the shape of the meniscus around two static fibers and the resulting capillary force. Note that this approach is, a priori, valid only for small meniscus slopes ($\theta_Y \simeq \pi/2$). However, Eq.~(\ref{eq-h0}) coincide with the expression of $h_{\text{s}}$ obtained with this approach and Fig.~\ref{fig3}(e) shows that it gives a good description of the data even for $\theta_Y=0$. In addition, we have shown that the capillary force obtained in Ref.~\cite{kralchevsky1992capillary} gives a good description of the coalescence of two static fibers when $\theta_Y=0$~\cite{siefert2022capillary}. In the following, we thus use $\theta_Y=0$. In this case, the difference of surface energy, $U_S$, between a configuration where two vertical fibers of radius $R$ are at an infinite distance and the one where they are at a distance $2d$ (Fig.~\ref{fig1}(a)) reads as
\begin{equation}
\label{energy}
    U_S=- 2\pi \gamma R\ (H_{d} - H_\infty),
\end{equation}
where $H_{d}$ and $H_\infty$ are the mean elevations of the (static) meniscus contact line on each fiber when they are, respectively, at a distance $2d$ or at infinity. When $d\gg R$, this difference is given by~\cite{kralchevsky1992capillary,siefert2022capillary}
\begin{equation}
H_{d} - H_\infty = R\ln[\ell_c/(2d\gamma_e)]=h_{\text{s}}/2-2R\ln 2.
\end{equation}
The static capillary force is given by $2F_{\text{s}} = -\partial U_S/\partial d$:
\begin{equation}
\label{F0}
F_{\text{s}}(d)=\frac{\pi}{2} \gamma R\, \frac{\partial h_{\text{s}}}{\partial d}=-\frac{\pi \gamma R^2}{d},
\end{equation}
where we used the logarithmic approximation of $h_{\text{s}}$, see Eq.~(\ref{eq-h0}). In both the static and the dynamical cases, the capillary force is mediated by the meniscus, therefore we assume that the interaction energy~\eqref{energy} and the method to derive the capillary force still apply with the static meniscus replaced by dynamical one. In this case, we obtain
\begin{equation}
\label{forcefinal}
    F(d)=\frac{\pi}{2} \gamma R\, \frac{\partial h}{\partial d}= (1+ \zeta) F_{\text{s}}(d), \quad \zeta \simeq 12.2\, \text{Ca}^{4/5},
\end{equation}
where the expression (\ref{eq-h}) of $h$ has been used.

\begin{figure}[t]
\centering
\includegraphics[width=\columnwidth]{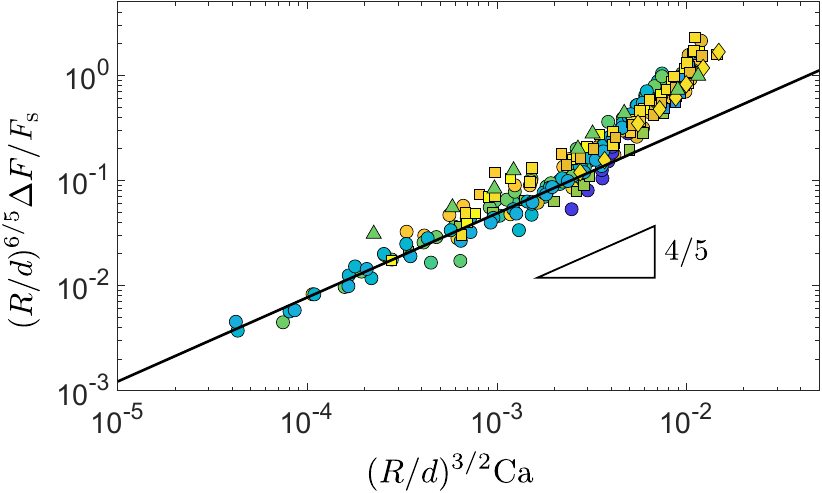}
\caption{Evolution of $\Delta F/F_{\text{s}}$ as a function of $\text{Ca}$. All data collapse on the power law obtained in Fig.~\ref{fig2}(d): $\Delta F/F_{\text{s}} = 12.2\, \text{Ca}^{4/5}$. An increase of slope is observed when $d\simeq 50 R\, \text{Ca}^{2/3}$. See Fig.~\ref{fig1} for the symbol and color code.}
\label{fig4}
\end{figure}

Figure~\ref{fig4} shows that the relative variation of the dynamical capillary force, $\Delta F/F_{\text{s}}$, computed from the raw data reported in Fig.~\ref{fig1}(d), varies in good approximation as $\zeta$ in agreement with Eq.~(\ref{forcefinal}). Note that the data in Fig.~\ref{fig4} has been again further rescaled by powers of $R/d$, as in Fig.~\ref{fig3}(d), to capture the change of slope as explained above. This change of slope occurs at $(R/d)^{3/2}\text{Ca}\simeq 3 \times 10^{-3}$, i.e. a value somewhat smaller than for $\Delta h/h_{\text{s}}$, see Figs.~\ref{fig3}(d) and \ref{fig4}. We attribute this difference to the fact that, in the force measurement experiments, the elastic fibers are deflected towards each other to typically half their initial distance $2d$ before snapping to contact~\cite{siefert2022capillary}.

Equation~(\ref{forcefinal}) shows that dynamical interaction between two fibers withdrawn from a liquid bath at a finite speed is equivalent to the static interaction with an effective surface tension increasing with the capillary number, i.e. $\gamma_{\text{eff}} = \gamma \, (1+\zeta)$. It appears thus that the potential hydrodynamic effects in the bath are negligible, as in the case of fibers sedimentation~\cite{guazzelli2011fluctuations}. Consequently, all the analysis performed for the static case can be easily reproduced~\cite{siefert2022capillary}. For example, neglecting tension in the fibers and assuming $d \gg R$ to keep the algebra simple here, the critical fiber length beyond which coalescence occurs during the withdrawal is given by
\begin{equation}
L^3 = \frac{L_{\text{s}}^3}{1+\zeta}=\frac{3B\, d^2}{4\pi \gamma \,R^2}(1+\zeta)^{-1}.
\end{equation}
The critical fiber length is thus reduced by 25\% compared to the static one at $\text{Ca} \simeq 0.065$ and by 50\% at $\text{Ca} \simeq 0.5$.

In the case of partially wetting liquids ($\theta_Y>0$), the increase in force is even more spectacular. For example, in the extreme $\theta_Y=\pi/2$, the interaction force vanishes in the static case, as the air-liquid interface is not deformed by the presence of the fibers. At finite retraction speed however, the viscous entrainment along the fibers leads to the deformation of the interface and to a new dynamic contact angle, leading to a finite capillary force~\cite{chan2012theory,shing2011maximum}.

In summary, we have studied experimentally and theoretically the withdrawal of two fibers out of a liquid bath. To account for the steep increase of the capillary attractive force, we first showed that the shape of the dynamical meniscus around one fiber can be obtained by scaling the shape of the static meniscus by a factor depending on Ca. We then experimentally demonstrated that the air-liquid interface between two withdrawn fibers is in good approximation described by the linear superposition of the interface around one fiber. With these ingredients, we derived analytical expressions for the evolution of the attraction force with Ca and the critical fiber length beyond which two fibers coalesce at a given Ca. A natural prolongation of this work is to consider an array of fibers to study potential collective effects on the onset of coalescence and on the shape of the coalesced states, i.e. single or multiple coalesced bundle of fibers. Indeed, recent experiments show that retraction speed
strongly influence the coalesced shape~\cite{tawfick2019,tawfick2020}. The influence of coalescence on the amount of fluid stored in an array of fibers is also of interest to optimize fluid capture with potential applications to rationalize the nectar feeding by some passerine birds characterized by brush-like tongue~\cite{Mitchell1990,Chang2013}. Finally, we note that the retraction speed appears to be an interesting parameter to fine tune the capillary interaction force between slender structures. It may open new possibilities for the capillary self-assembly of particles and fibers at fluid interfaces~\cite{pokroy2009self,zeng20223d}.

\begin{acknowledgments}
The authors acknowledge support by F.R.S.-FNRS under the research grant (PDR ``ElastoCap'') n$^{\circ}$ T.0025.19 and n$^{\circ}$ J.0017.21 (CDR ``FASTER''). This project has received funding from the European Union's Horizon 2020 research and innovation programme under the Marie Sklodowska-Curie grant agreement n$^{\circ}$ 101027862. This project also has received the support of ULB incentives measures, available thanks to the FWB, to foster participation to EU projects.
\end{acknowledgments}

H.B. and E.S. contributed equally to this work.

\clearpage

\appendix

{\Large \textbf{Supplemental Material}}

\section{Experimental methods}

We consider a pair of identical fibers partially immersed in a liquid. The liquid container is large compared to the spatial extension of the menisci to avoid boundary effects. The cylindrical fibers, made of glass ($E=64$ GPa) or PET ($E=10$ GPa), have a radius $R$ ($50<R<160$ $\mu$m) and a length $L_0$ ($L_0\simeq 10$ cm). Silicon oil with surface tension $\gamma=0.021$ N/m and density $\rho_{\ell}=960$ kg/m$^{3}$ is used as a model fluid in most experiments. The kinematic viscosity $\nu$ is also varied ($100\le \nu \le 1000$ cSt, i.e. $0.096 \le \mu = \rho_{\ell} \nu \le 0.96$ Pa s). 
Glycerol is also used in one set of experiments to vary the surface tension of the liquid ($\gamma=0.063$ N/m, $\rho_{\ell}=1260$ kg/m$^{3}$). As the viscosity of glycerol is highly temperature and humidity dependent, it is measured experimentally before and after the set of experiments by dropping a small metallic bead inside the container and measuring its terminal velocity. We measure a slight drop of viscosity during a set of experiments ($\mu=1.33$ Pa s before and $\mu=1.27$ Pa s after the experiments). The mean viscosity value is used for this set of experiments. Note that, as glycerol does not wet glass, plasma treated glass fibers have been used for this set of experiments.

\begin{figure}[!t]
\centering
\includegraphics[width=\columnwidth]{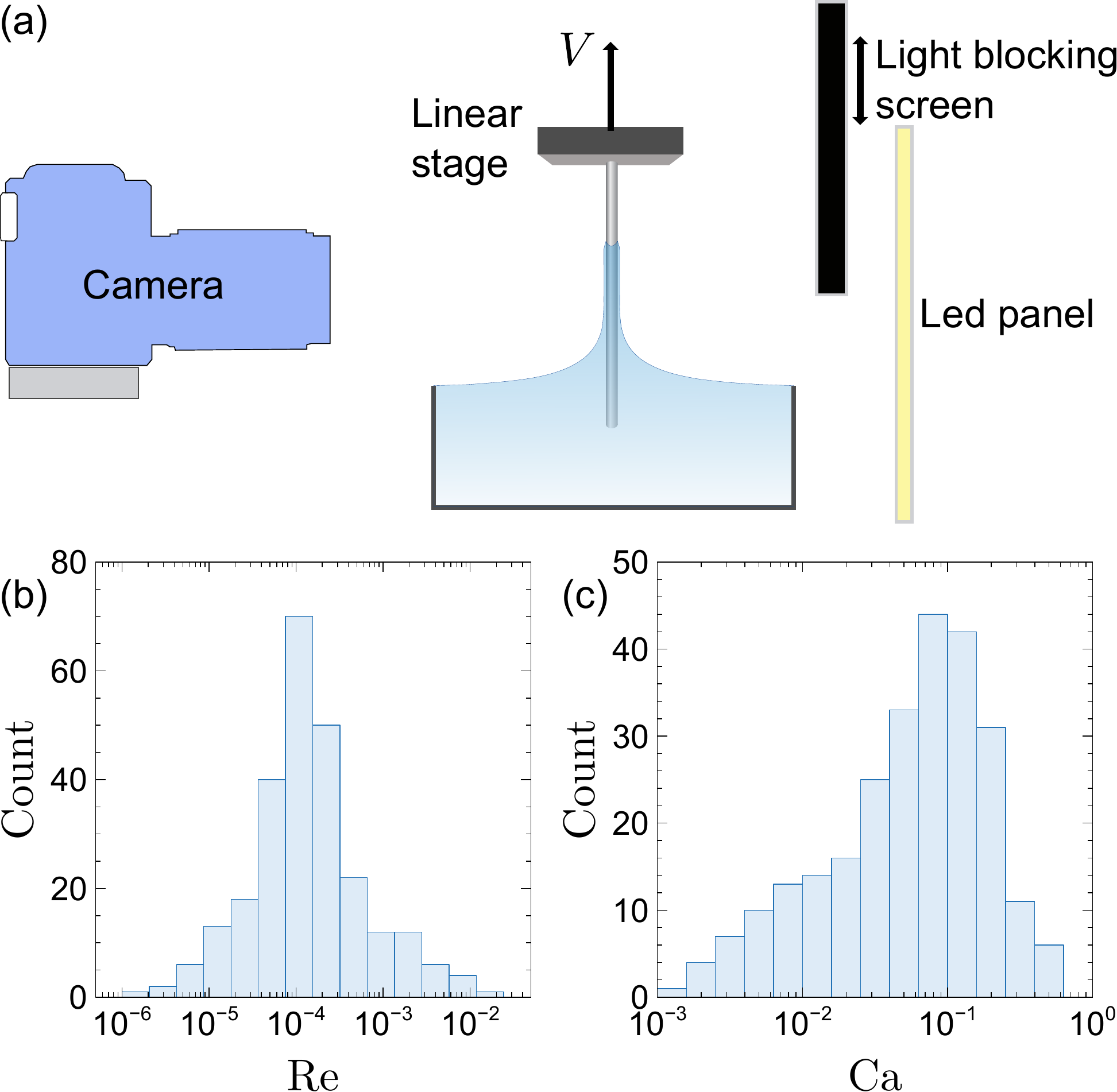}
\caption{(a) Schematics of the set up: a fiber is withdrawn from a fluid bath at a controlled speed $V$ with a linear stage and filmed with a camera. The scene is illuminated with a led panel, partially obstructed by an adjustable screen. The shallow light setting allows for an enhancement of the contrast: small slopes of the fluid interface do not transmit light and the meniscus thus appears black on the camera. Distribution of the Reynolds number (b), $\text{Re} = \rho_{\ell} V R/\mu$, and of the capillary number (c), $\text{Ca} = \mu V/\gamma$, considered in the experiments.}
\label{figSI:setup}
\end{figure}

For all experiments, fibers are clamped vertically to a traction machine (ZwickiLine Z0.5 from Zwick), at a distance $2d$ ($0.2<d<1.3$ mm). The initial gap is small with respect to their dry length, $d/L\ll 1$. Initially, the fibers are immersed in the liquid deep enough such that they do not coalesce. Therefore, the distance between the two structures at the interface, $2\delta$, is close to the imposed distance at the clamped ends, $2d$; i.e. the structures are quasi-parallel. They are quasi-statically withdrawn by steps of typically 1~mm with a typical speed of 2 mm/min, until coalescence is reached for a critical dry length $L_{\text{s}}$. When $d$ is too large, the fibers may not coalesce during a quasi-static withdrawal, whereas they will do beyond a given finite retraction speed, because the fiber length $L_0$ is smaller than $L_{\text{s}}$. In such a case, $L_{\text{s}}$ cannot be measured and its value is computed from the theory developed in Ref.~\cite{siefert2022capillary}.
The two fibers are then cut to a smaller length $L<L_{\text{s}}$, such that they do no longer quasi-statically coalesce. The retraction speed $V$ is then gradually increased in order to find the smallest one $V_c$ for which the two fibers coalesce. In all experiments, $1.6\times 10^{-6} \le \text{Re} \le 0.012$ and $1.2\times 10^{-3} \le \text{Ca} \le 0.5$, see Fig.~\ref{figSI:setup}(b),(c).
Note that to account for the potential slight deviation from parallelism between both structures (due to misalignment or a small natural curvature of the fibers), the distance $2d$ between both structures is measured after they are completely removed from the bath at the free end of the fibers.

In order to observe the menisci around one or two fibers, we fill up the container with liquid (silicon oil or glycerol) up to the top edge, in order to get rid of the meniscus on the lateral walls. We illuminate the experimental set up using a LED panel placed behind the bath, partially obstructed by an adjustable screen, see Fig.~\ref{figSI:setup}(a). This shallow backlighting setup increases the contrast: small slopes of the fluid interface deflect the light and, as a consequence, appear black on the pictures (as in Figs.~2(a) and 3(a) of the main text).

\section{Theoretical model}

\subsection{General equations and numerical method}

\begin{figure}[t]
\centering
\includegraphics[width=\columnwidth]{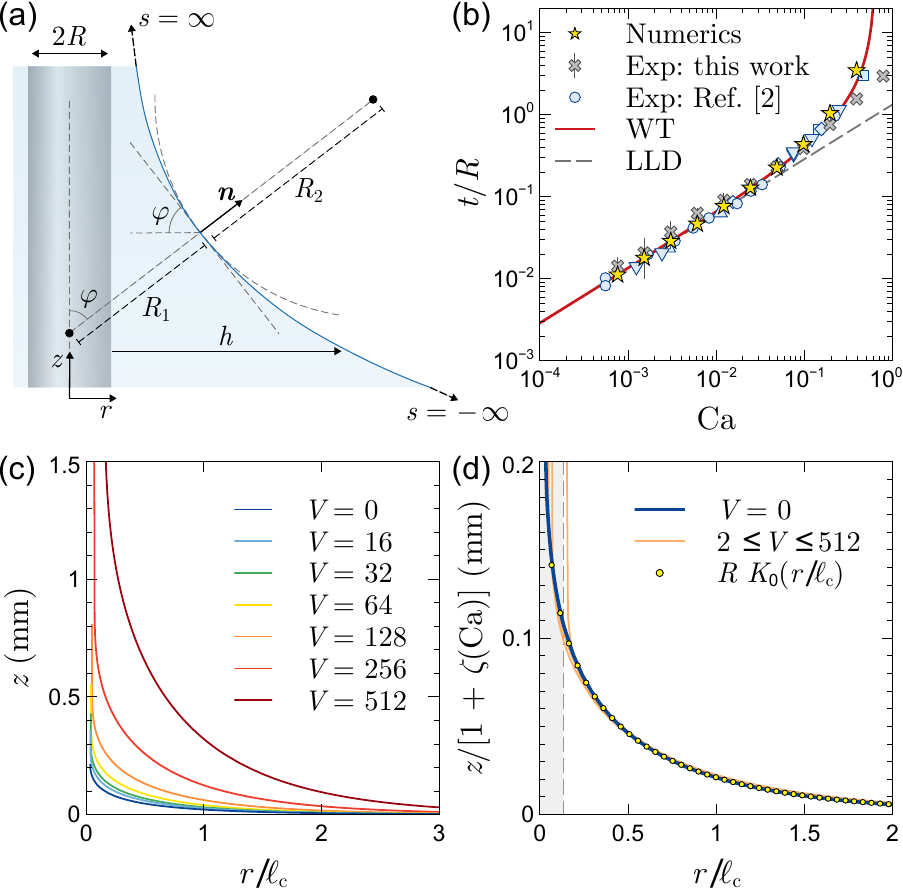}
\caption{(a) Schematic of the system showing the arclength $s$, the angle $\varphi$ along the interface and the two principal radii of curvature at a given point ($\kappa_1 = R_1^{-1} = \sin \varphi/(R+h)$, $\kappa_2=R_2^{-1} = -d\varphi/ds$). (b) Comparison between the experimental, theoretical and numerical evolution of the liquid layer thickness as a function of the capillary number. The numerical data are obtained by solving Eqs.~(\ref{model-adim}). The solid and dashed curves correspond to Eqs.~(\ref{t-LLD-app}) and (\ref{t-WT}). Grey cross: glass fiber ($R=100$ $\mu$m), $\mu = 0.96$ Pa s, $\gamma = 21$ mN/m. The blue symbols refers to data from Ref.~\cite{quere96} obtained using a nickel wire ($R=63.5$ $\mu$m) and a tungsten wire ($R=12.5$ $\mu$m) and a withdrawal velocity varying between 0.15 mm/s and 50 mm/s. Circle: $\mu = 0.019$ Pa s, $\gamma = 20.6$ mN/m; Diamond: $\mu = 0.096$ Pa s, $\gamma = 20.9$ mN/m; Square: $\mu = 0.291$ Pa s, $\gamma = 21.1$ mN/m; Triangle down: $\mu = 0.485$ Pa s, $\gamma = 21.1$ mN/m; Triangle up: $\mu = 12.250$ Pa s, $\gamma = 21.1$ mN/m. (c) Meniscus shape as a function of the rescaled radial distance $r/\ell_c$ at various $V$ (mm/min) obtained by solving numerically Eqs.~(\ref{model-adim}) with $\mu = 0.96$ Pa s, $\gamma = 21$ mN/m and $\text{Bo}=1/900$. (d) Collapse of the dynamical meniscus profiles onto the static meniscus shape when their height are properly rescaled by $1+\zeta$. $V=2^n$ mm/min with $1 \le n \le 9$.}
\label{figSI:theory}
\end{figure}

The model proposed in Ref.~\cite{zhang2022dip} has been used to compute the thickness $t$ of the liquid layer and the meniscus profile. This model reads as
\begin{subequations}
\label{model-dim}
\begin{align}
\label{z-h-phi}
\frac{dz}{ds}&=\sin \varphi, \quad \frac{dh}{ds}=-\cos \varphi, \quad \frac{d\varphi}{ds}=-\frac{p}{\gamma}+\frac{\sin \varphi}{R+h}, \\
\label{p}
\frac{dp}{ds}&= \left[3 \mu V \left(\frac{h-h_{\infty}}{h^3}\right)- \rho_{\ell} g \left(1-\frac{h_{\infty}^3}{h^3}\right) \right] \sin \varphi,
\end{align}
\end{subequations}
where $-\infty < s < \infty$ is the arclength of the air-liquid interface, $\varphi$ the local angle between the tangent to the interface and the horizontal axis (see Fig.~\ref{figSI:theory}(a)), $h_{\infty} \equiv t$ the thickness of the liquid layer when $s\to \infty$, $p$ the pressure in the liquid and $g$ the gravitational acceleration (the other symbols have been defined above). The first two of Eqs.~(\ref{z-h-phi}) are the geometric relationships between, $z$, $h$, $\varphi$ and $s$. The third of Eqs.~(\ref{z-h-phi}) is the Laplace pressure $p=\gamma (\kappa_1 + \kappa_2)$. Equation~(\ref{p}) is obtained from a standard lubrication approximation assuming that the system is in a steady state.

Rescaling all lengths by $R$, the pressure by $\gamma/R$, using capital letter for dimensionless variables and using $h$ as independent variables, we have
\begin{subequations}
\label{model-adim}
\begin{align}
\label{z-h-phi-adim}
\frac{dZ}{dH}&=-\tan \varphi, \quad \cos \varphi \frac{d\varphi}{dH}=P-\frac{\sin \varphi}{1+H}, \\
\label{p-adim}
\frac{dP}{dH} &= \left[-3 \text{Ca} \left[\frac{H-H_{\infty}}{H^3}\right]+ \text{Bo} \left[1-\frac{H_{\infty}^3}{H^3}\right] \right] \tan \varphi,
\end{align}
\end{subequations}
where $\text{Bo} = R^2/\ell_c^2$ is the Bond number. The equations for $\varphi$ and $P$ are integrated using the initial conditions
\begin{equation}
\label{cond-ini-P-phi}
P(H_{\infty}) = \frac{1}{1+H_{\infty}}, \quad \varphi(H_{\infty}) = \frac{\pi}{2}.
\end{equation}
To fix the unknown quantity $H_{\infty}$, an additional condition is needed and is obtained from the asymptotic analysis of Eqs.~(\ref{model-adim}). When $H\to \infty$, $\varphi \ll 1$ and we have asymptotically
\begin{equation}
\label{phi-P-asym}
\frac{dZ}{dH}=-\varphi, \quad \frac{d\varphi}{dH}=P-\frac{\varphi}{H}, \quad \frac{dP}{dH}= \text{Bo}\, \varphi.
\end{equation}
Eliminating $P$, we get
\begin{equation}
\label{phi-asymp}
\frac{d}{dH}\left(\frac{d\varphi}{dH} + \frac{\varphi}{H}\right)= \text{Bo}\, \varphi, \quad \Rightarrow \quad \varphi = \bar{c}\, K_1(\sqrt{\text{Bo}}\, H),
\end{equation}
the other independent solution diverges when $H\to \infty$ and must be discarded. Using the second of Eqs.~(\ref{phi-P-asym}), we obtain
\begin{equation}
P = - \bar{c}\, \sqrt{\text{Bo}}\, K_0(\sqrt{\text{Bo}}\, H).
\end{equation}
Therefore, we have
\begin{equation}
\label{cond-asymp-P-phi}
\frac{P}{\varphi} = -\frac{\sqrt{\text{Bo}}\, K_0(\sqrt{\text{Bo}}\, H)}{K_1(\sqrt{\text{Bo}}\, H)} \quad \text{as} \quad H\to \infty,
\end{equation}
which does no longer depend on the arbitrary constant $\bar{c}$. Note that, using the first of Eqs.~(\ref{phi-P-asym}) together with the expression~(\ref{phi-asymp}) of $\varphi$, we have
\begin{equation}
\label{Z-asymp}
Z = c\, K_0(\sqrt{\text{Bo}}\, H) \quad \text{as} \quad H\to \infty.
\end{equation}

In practice, Eqs.~(\ref{model-adim}) for $\varphi$ and $P$ are integrated with Mathematica between $H=H_{\infty}$ and $H=H_{\text{max}}$ using the initial conditions (\ref{cond-ini-P-phi}). $H_{\infty}$ is varied until the condition (\ref{cond-asymp-P-phi}) is satisfied at $H=H_{\text{max}}$. We used $300 \le H_{\text{max}} \le 415$ depending on the value of $\text{Ca}$ so that $\varphi(H_{\text{max}}) \simeq 10^{-7}$. Finally, once $\varphi$ is known, the first of Eqs.~(\ref{z-h-phi-adim}) is integrated with the boundary condition $Z(H_{\text{max}}) = 0$.

\subsection{Numerical solutions}

Figure~\ref{figSI:theory}(b)-(d) shows some numerical solutions obtained by solving Eqs.~(\ref{model-adim}) with the conditions (\ref{cond-ini-P-phi}) and (\ref{cond-asymp-P-phi}). Figure~\ref{figSI:theory}(b) shows the evolution of $H_{\infty}\equiv t/R$ as a function of $\text{Ca}$ for $\text{Bo}=1/900$ ($R=0.05$ mm and $\ell_c = 1.5$ mm). This evolution agrees with the Landau-Levich-Derjaguin (LLD) relation
\begin{equation}
\label{t-LLD-app}
   t=1.34\, R\, \text{Ca}^{2/3},
\end{equation}
up to $\text{Ca}\simeq 10^{-2}$ and with the White-Tallmadge (WT) extension~\cite{white1966theory}
\begin{equation}
\label{t-WT}
   t=1.34\, R\, \text{Ca}^{2/3} \left[1- 1.34\, \text{Ca}^{2/3}\right]^{-1},
\end{equation}
up to $\text{Ca}\simeq 0.5$. Both numerical and theoretical results agree well with experimental data also up to $\text{Ca}\simeq 0.5$.

\begin{figure}[t]
\centering
\includegraphics[width=\columnwidth]{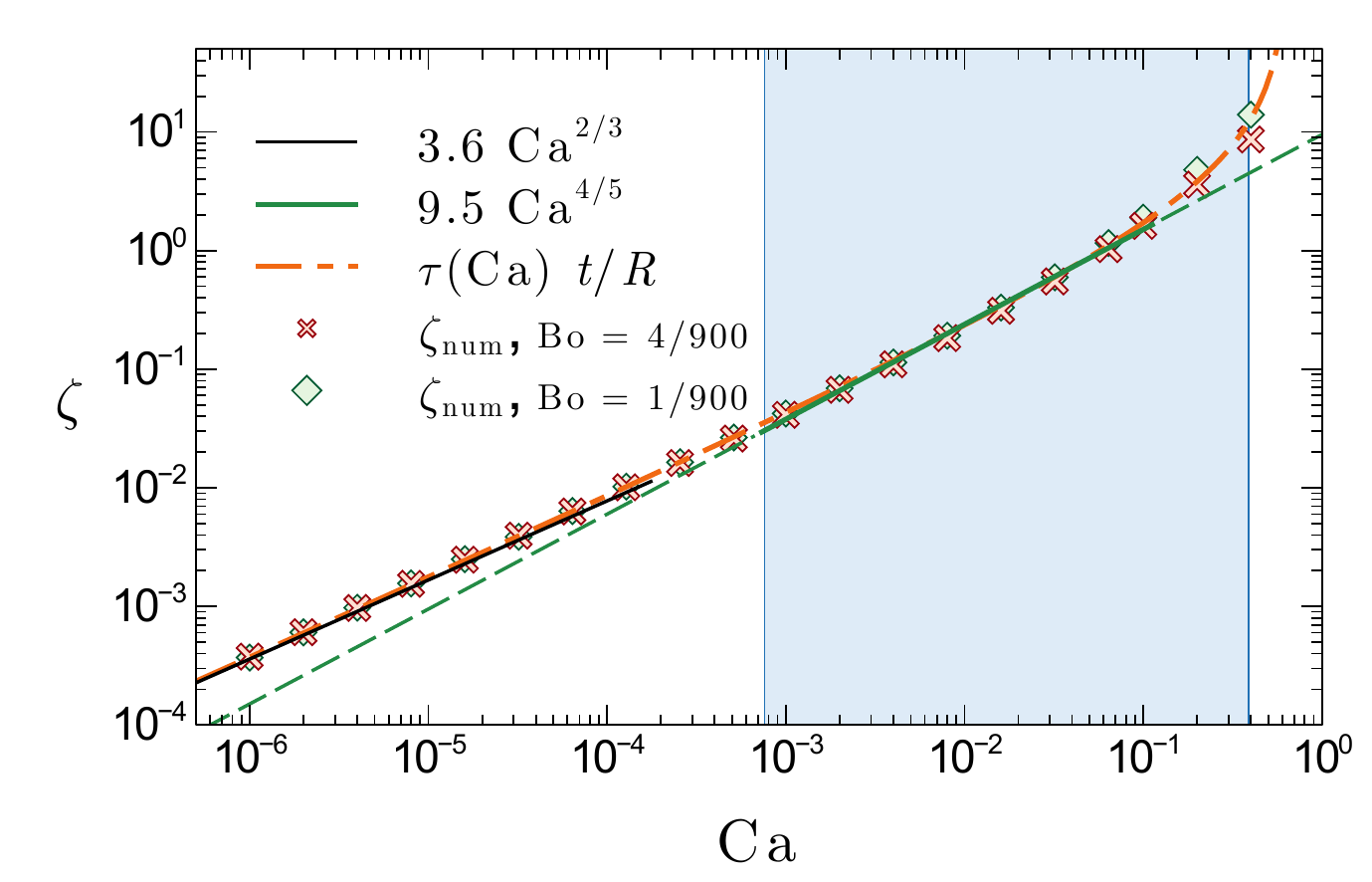}
\caption{Evolution of the scaling factor $\zeta_{\text{num}}$ as a function of the capillary number showing its approximate scaling at low and intermediate values of $\text{Ca}$. The evolution is shown for two values of the Bond number related to the experiments ($\ell_c = 1.5$ mm, $R=50$ and $100$ $\mu$m). $\zeta_{\text{num}}$ follows closely the evolution of $t$ given by Eq.~(\ref{t-WT}) multiplied by a function $\tau$ varying slowly with $\text{Ca}$, $\tau = 2.7 (1+ \text{Ca}^{1/4})$. The blue shaded area correspond to the interval of $\text{Ca}$ considered in this work.}
\label{figSI:zeta}
\end{figure}

Figure~\ref{figSI:theory}(c) shows the meniscus profile for various retraction speed ($1.5 \times 10^{-3} \le \text{Ca} \le 0.39$). Figure~\ref{figSI:theory}(d) shows that they collapse onto the static meniscus shape when their height are properly rescaled by $1+\zeta_{\text{num}}$ as for the experimental profiles (see Fig.~2 of the main text). The evolution of the scaling factor $\zeta_{\text{num}}$ as a function of $\text{Ca}$ is shown in Fig.~\ref{figSI:zeta}. In the interval of $\text{Ca}$ considered in this work, $\zeta_{\text{num}} \sim \text{Ca}^{4/5}$ as reported in the main text. However, for $\text{Ca}\lesssim 10^{-4}$, the scaling slightly changes to $\zeta \sim \text{Ca}^{2/3}$. In this regime, the dynamical capillary force is still given by $F=(1+ \zeta) F_{\text{s}}$, where $F_{\text{s}}$ is the static capillary force, but with $\zeta\sim \text{Ca}^{2/3}$ instead of $\text{Ca}^{4/5}$.

\subsection{Static case}

In the static case, $\text{Ca}=H_{\infty}=0$ and Eqs.~(\ref{model-adim}) reduce to
\begin{equation}
\label{static-temp}
Z'=-\tan \varphi, \ (\sin \varphi)'=P-\frac{\sin \varphi}{1+H}, \ \frac{P'}{\text{Bo}}=  \tan \varphi,
\end{equation}
where prime denotes a $H$ derivative. From the first and third of Eqs.~(\ref{static-temp}), we get
\begin{equation}
\label{P-temp}
P = - \text{Bo}\, Z,
\end{equation}
where we used $P(0)=0$ to fix the integration constant. From the first of Eqs.~(\ref{static-temp}), we get
\begin{equation}
\label{sin-phi}
\sin \varphi = -\frac{Z'}{\sqrt{1+Z'^2}}.
\end{equation}
Substituting Eqs.~(\ref{P-temp}) and (\ref{sin-phi}) in the second of Eqs.~(\ref{static-temp}), we finally have
\begin{equation}
\label{eq-static}
\text{Bo}\, Z = \frac{Z''}{(1+Z'^2)^{3/2}} + \frac{Z'}{(1+H)\sqrt{1+Z'^2}}.
\end{equation}
This is the standard equation for a static meniscus balancing the hydrostatic pressure (left) with the Laplace pressure (right), see for example~\cite{Lo1983}. 

Far enough from the fiber, i.e. $H\gg 1$, we have $Z' \ll 1$ and the last equation reduces to $\text{Bo}\, Z_{\text{s},\text{o}} = Z_{\text{s},\text{o}}'' + Z_{\text{s},\text{o}}'/H$, whose general solution reads
\begin{equation}
Z_{\text{s},\text{o}} = c\, K_0(\sqrt{\text{Bo}}H) + \bar{c} \, I_0(\sqrt{\text{Bo}}H),
\end{equation}
where $K_0$ and $I_0$ are the two modified Bessel functions of zeroth order~\cite{NIST} and where the indexes $\text{s}$ and $\text{o}$ stands for static and outer respectively. Requiring that $Z \to 0$ as $H$ diverges leads to $\bar{c}=0$. The other constant of integration is fixed by matching the outer solution, $Z_{\text{o}}$, with the inner solution, $Z_{\text{i}}$, valid near the fiber, i.e. $H\lesssim 1$. In this region, the second curvature term in the right-hand side of Eq.~(\ref{eq-static}) necessarily tends to 1 ($1/R$ in physical units) when $H\to 0$ so that the left-hand side term in Eq.~(\ref{eq-static}) is negligible for small Bond number. Equation~(\ref{eq-static}) reduces to $Z_{\text{s},\text{i}}'' + Z_{\text{s},\text{i}}'(1+Z_{\text{s},\text{i}}'^2)/(1+H)=0$. Imposing that $Z_{\text{s},\text{i}}' \to -\infty$ as $H\to 0$ (vanishing contact angle), we get
\begin{equation}
\label{eq-static-inner}
Z_{\text{s},\text{i}} = \tilde{c} -\ln\left[1+H + \sqrt{(1+H)^2-1}\right],
\end{equation}
where the integration constant $\tilde{c}$ is simply the height of the meniscus on the fiber (see Ref.~\cite{Lo1983} for a more general treatment valid for any contact angle). The matching is done by requiring that
\begin{equation}
Z_{\text{match}} = \lim_{H\to \infty} Z_{\text{s},\text{i}} = \lim_{H\to 0} Z_{\text{s},\text{o}}.
\end{equation}
We have
\begin{subequations}
\begin{align}
\lim_{H\to \infty} Z_{\text{s},\text{i}} &= \tilde{c}- \ln 2-\ln H, \\
\lim_{H\to 0} Z_{\text{s},\text{o}} &= -c \ln(\gamma_e \sqrt{\text{Bo}}/2) - c \ln H,
\end{align}
\end{subequations}
where $\gamma_e \simeq 1.781$ is the exponential of the Euler-Mascheroni constant. We get $c=1$ and $\tilde{c}=\ln(4/\gamma_e \sqrt{\text{Bo}})$. We thus obtain, in physical units
\begin{subequations}
\begin{align}
\label{eq-static-inner-final}
z_{\text{s},\text{i}} &= R \ln\left[\frac{4}{\gamma_e \sqrt{\text{Bo}}}\right] - R \ln\left[\frac{r}{R} + \sqrt{\frac{r^2}{R^2}-1}\right], \\
\label{eq-static-outer-final}
z_{\text{s},\text{o}} &= R\, K_0(r/\ell_c), \\
\label{eq-static-match-final}
z_{\text{match}} &= R \ln(2/\gamma_e \sqrt{\text{Bo}}) - R \ln(r/R).
\end{align}
\end{subequations}
The composite solution, valid over the interval $R \le r < \infty$, is obtained from $z = z_{\text{s},\text{i}} + z_{\text{s},\text{o}} - z_{\text{match}}$ and reads as
\begin{equation}
\frac{z_{\text{s}}(r)}{R}= - \ln\left[\frac{r + \sqrt{r^2-R^2}}{2r}\right] + K_0\left(\frac{r}{\ell_c}\right).
\end{equation}

\subsection{Dynamic case}

As shown in Eq.~(\ref{Z-asymp}), the meniscus profile far from the fiber behaves as in the static case
\begin{equation}
\label{eq-dynamic-outer}
Z_{\text{d},\text{o}} = c\, K_0(\sqrt{\text{Bo}}H),
\end{equation}
where the constant $c$ must be fixed by matching with the inner solution near the fiber and the index $\text{d}$ stands for dynamic. To obtain the inner solution, it is easier to use $Z$ as the independent variable so that Eqs.~(\ref{model-adim}) become
\begin{subequations}
\label{model-adim2}
\begin{align}
\label{z-h-phi-adim2}
\frac{dH}{dZ}&=-\cot \varphi, \quad \sin \varphi \frac{d\varphi}{dZ}=-P+\frac{\sin \varphi}{1+H}, \\
\label{p-adim2}
\frac{dP}{dZ}&= 3 \text{Ca} \left(\frac{H-H_{\infty}}{H^3}\right)- \text{Bo} \left(1-\frac{H_{\infty}^3}{H^3}\right).
\end{align}
\end{subequations}
To describe the meniscus shape near the fiber, we write
\begin{subequations}
\label{expansion}
\begin{align}
\label{expansion-H-phi}
H &=H_{\infty} + \epsilon\, H_1 + \epsilon^2\, H_2, \quad \varphi = \frac{\pi}{2}- \epsilon\, \varphi_1 - \epsilon^2\, \varphi_2, \\
P &=\frac{1}{1+H_{\infty}}- \epsilon\, P_1 - \epsilon^2\, P_2.
\end{align}
\end{subequations}
Those expansions are used in Eqs.~(\ref{model-adim2}) which are solved order by order. They are obviously satisfied at order $\epsilon^0$.

\subsubsection{Order $\epsilon$}

At order $\epsilon$, we get the system of equations
\begin{subequations}
\begin{align}
H_1' &=-\varphi_1, \quad \varphi_1'=-P_1 +\frac{H_1}{(1+H_{\infty})^2}, \\
P_1' &=-3\left(\frac{\text{Ca}}{H_{\infty}^3}-\frac{\text{Bo}}{H_{\infty}}\right)H_1.
\end{align}
\end{subequations}
In the following, we consider $\text{Ca}\gg \text{Bo}H_{\infty}^2$ which somewhat simplifies the last equation. Eliminating $\varphi_1$ and $P_1$, we get
\begin{equation}
H_1''' + \frac{H_1'}{(1+H_{\infty})^2}+\frac{3\,\text{Ca}}{H_{\infty}^3}H_1=0.
\end{equation}
The solution reads $H_1 = \tilde{c}_1\, \exp(-\sigma\, Z)$ with $\sigma$ solution of
\begin{equation}
\label{eq-sigma}
\sigma^3 + \frac{\sigma}{(1+H_{\infty})^2} -\frac{3\,\text{Ca}}{H_{\infty}^3}=0, \quad \Rightarrow\quad \sigma \simeq \frac{(3\,\text{Ca})^{1/3}}{H_{\infty}},
\end{equation}
because the last term of the equation to solve diverges when $\text{Ca} \to 0$. Therefore, we get
\begin{equation}
\label{H1-phi1-P1}
H_1 = \tilde{c}_1 e^{-\sigma Z}, \quad \varphi_1= \tilde{c}_1 \sigma e^{-\sigma Z}, \quad P_1 = \tilde{c}_1 \sigma^2 e^{-\sigma Z},
\end{equation}
where we took into account the fact that $\sigma$ diverges when $\text{Ca} \to 0$ in the expression of $\varphi_1$. At order $\epsilon$, we thus get
\begin{equation}
H = H_{\infty} + c_1 e^{-\sigma Z},
\end{equation}
where $c_1 = \epsilon \tilde{c}_1$. Inverting this last relation, we get
\begin{equation}
\label{eq-dynamic-inner-1}
Z_{\text{d},\text{i}} = Z_0 -\frac{1}{\sigma}\ln(H - H_{\infty}),
\end{equation}
where $Z_0$ is so far arbitrary. This solution is now matched with the inner static solution (\ref{eq-static-inner}) by imposing the equality at some $H^{\star}$ between the two profiles and their two first derivatives. However, there is only two free parameters, $H^{\star}$ and $Z_0$, for three conditions. Nevertheless, the static meniscus does no longer connect to the fiber of radius $R$ but instead to an effective fiber with a radius $\mathcal{R} \ge R + h_0$. Therefore, Eq.~(\ref{eq-dynamic-inner-1}) is instead matched to
\begin{equation}
\label{eq-static-inner-bis}
\bar{Z}_{\text{s},\text{i}} = \bar{R}\left\{\bar{c} -\ln\left[\frac{1+H}{\bar{R}} + \sqrt{\frac{(1+H)^2}{\bar{R}^2}-1}\right]\right\},
\end{equation}
where $\bar{R}=\mathcal{R}/R$ and where $\bar{c}$ will be fixed by matching the outer solution. Imposing at $H=H^{\star}$ that
\begin{equation}
Z_{\text{d},\text{i}}= \bar{Z}_{\text{s},\text{i}}, \quad Z_{\text{d},\text{i}}'= \bar{Z}_{\text{s},\text{i}}', \quad Z_{\text{d},\text{i}}''= \bar{Z}_{\text{s},\text{i}}'',
\end{equation}
so that the two principal curvatures match, we get
\begin{align}
\label{barR-hs-z0}
\bar{R} &= \frac{\sqrt{1+\sigma^2(1+H_{\infty})^2}}{\sigma},\quad H^{\star} = H_{\infty} + \frac{1}{\sigma^2(1+H_{\infty})}, \nonumber \\
Z_0 &= \frac{\ln(H^{\star}-H_{\infty})}{\sigma}+\bar{Z}_{\text{s},\text{i}}(H^{\star}).
\end{align}
Using the expression (\ref{eq-sigma}) of $\sigma$ and $H_{\infty}= 1.34\, \text{Ca}^{2/3}$, we get at the lowest order in $\text{Ca}$
\begin{align}
\label{barR-hs-z0-2}
\bar{R} &\simeq 1+1.77\, \text{Ca}^{2/3},\quad H^{\star} \simeq 2.20\, \text{Ca}^{2/3}, \nonumber \\
Z_0 &\simeq \bar{c} + 0.62\, \ln(\text{Ca})\, \text{Ca}^{1/3}.
\end{align}
Finally, $\bar{Z}_{\text{s},\text{i}}$ is matched with the outer solution (\ref{eq-dynamic-outer}) by imposing
\begin{equation}
Z_{\text{match}} = \lim_{H\to \infty} \bar{Z}_{\text{s},\text{i}} = \lim_{H\to 0} Z_{\text{d},\text{o}}.
\end{equation}
We have
\begin{subequations}
\begin{align}
\lim_{H\to \infty} \bar{Z}_{\text{s},\text{i}} &= \bar{R} \bar{c}- \bar{R}\ln(2/\bar{R})- \bar{R} \ln H, \\
\lim_{H\to 0} Z_{\text{d},\text{o}} &= -c \ln(\gamma_e \sqrt{\text{Bo}}/2) - c \ln H.
\end{align}
\end{subequations}
Therefore, $c=\bar{R}$ and $\bar{c}=\ln(4/\gamma_e \sqrt{\text{Bo}}\, \bar{R})$. We thus get, in physical units
\begin{subequations}
\begin{align}
\label{eq-dynamic-inner-final-1}
\frac{\bar{z}_{\text{s},\text{i}}}{\mathcal{R}} &=  \ln\left[\frac{4}{\gamma_e \sqrt{\text{Bo}}\, \bar{R}}\right] - \ln\left[\frac{r}{\mathcal{R}} + \sqrt{\frac{r^2}{\mathcal{R}^2}-1}\right], \\
\label{eq-dynamic-outer-final-1}
\frac{z_{\text{d},\text{o}}}{\mathcal{R}} &= K_0(r/\ell_c), \\
\label{eq-dynamic-match-final-1}
\frac{z_{\text{match}}}{\mathcal{R}} &=  \ln(2/\gamma_e \sqrt{\text{Bo}}) - \ln(r/R),
\end{align}
\end{subequations}
where $\mathcal{R}= \bar{R}\,R \simeq R\, (1+1.77\, \text{Ca}^{2/3})$. The composite solution, valid over the interval $R+h^{\star} \le r < \infty$, is obtained from $z = \bar{z}_{\text{s},\text{i}} + z_{\text{d},\text{o}} - z_{\text{match}}$ and reads as
\begin{subequations}
\begin{align}
\frac{z_{\text{d}}^{(2)}(r)}{\mathcal{R}} &= - \ln\left[\frac{r + \sqrt{r^2-\mathcal{R}^2}}{2r}\right] + K_0\left(\frac{r}{\ell_c}\right),\\
\mathcal{R} &= \bar{R}\,R \simeq R\, \left(1+1.77\, \text{Ca}^{2/3}\right).
\end{align}
\end{subequations}
The solution valid over the interval $R+h_{\infty} \le r \le R+h^{\star}$ is obtained from Eq.~(\ref{eq-dynamic-inner-1}) 
\begin{equation}
\frac{z_{\text{d}}^{(1)}(r)}{R} = Z_0 -\frac{1}{\sigma}\ln\left[\frac{r-R - h_{\infty}}{R}\right]
\end{equation}
where $h^{\star} = R\, H^{\star}$ and $Z_0$ are given by Eqs.~(\ref{barR-hs-z0-2}). Finally, the solution valid over the interval $R+h_{\infty} \le r < \infty$ is given by
\begin{equation}
z_{\text{d}}(r)=z_{\text{d}}^{(1)}(r)\, \theta(R+h^{\star}-r) + z_{\text{d}}^{(2)}(r)\, \theta(r-R-h^{\star}),
\end{equation}
where $\theta(x)$ is the Heaviside function. 

Therefore, far enough from the fiber, the dynamic meniscus profile is given by $z_{\text{d}}\simeq \mathcal{R} K_0(r/\ell_c)$ whereas the static meniscus profile reads as $z_{\text{s}}\simeq R K_0(r/\ell_c)$. Consequently, their relative difference is given by $(\mathcal{R}-R)/R \simeq 1.77\, \text{Ca}^{2/3}$. We recover the scaling obtained numerically at small $\text{Ca}$, see Fig.~\ref{figSI:zeta}, yet with a smaller coefficient. The agreement can be slightly improved at the next order.

\begin{figure}[t]
\centering
\includegraphics[width=\columnwidth]{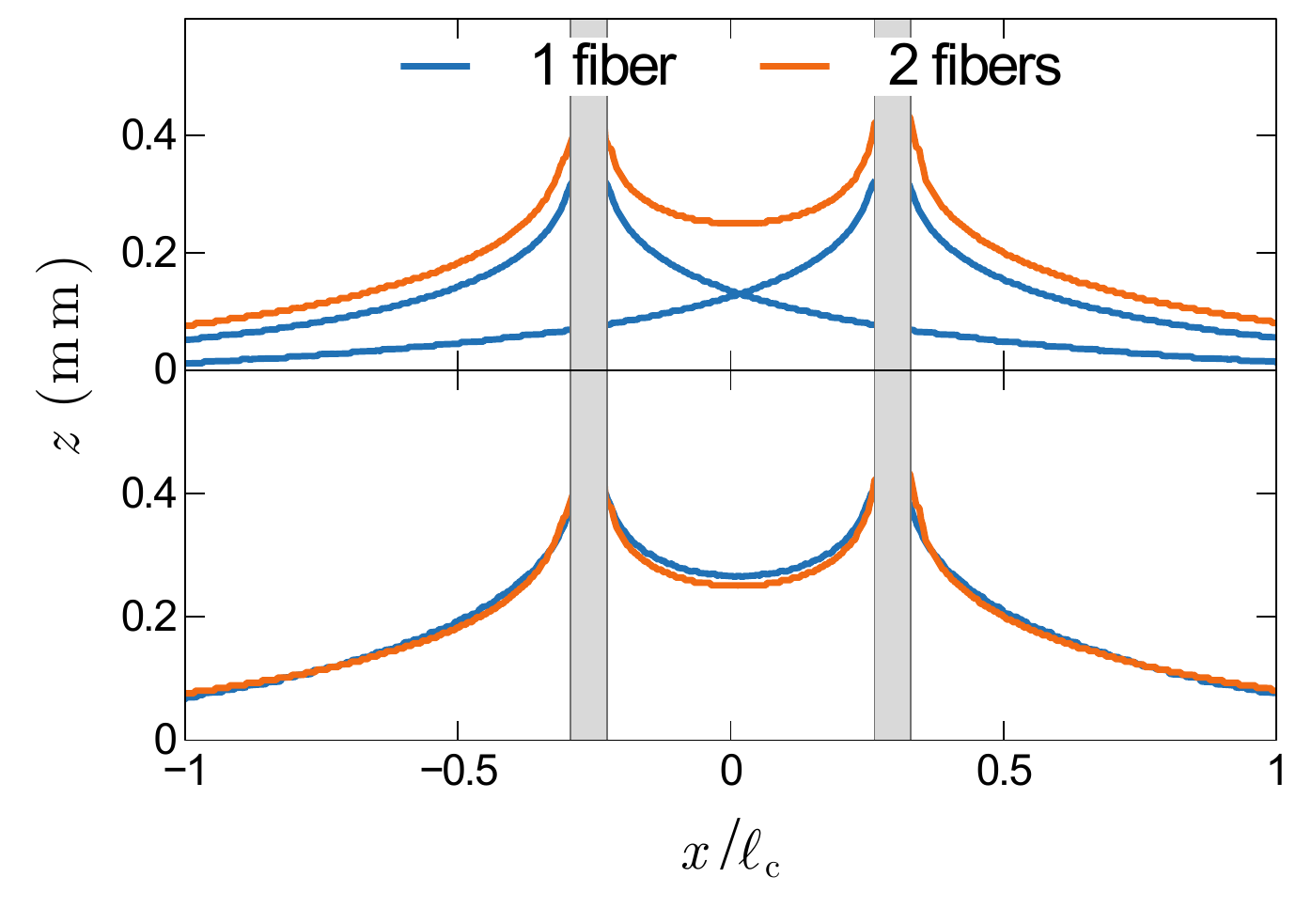}
\caption{Dynamical meniscus around two fibers ($R=50$ $\mu$m) separated by a distance $2d = 0.82$ mm and withdrawn from silicon oil at $V=32$ mm/min ($\text{Ca} =0.024$) reconstructed from the dynamical meniscus around a single fiber withdrawn at the same speed.}
\label{figSI:superpo}
\end{figure}

\subsubsection{Order $\epsilon^2$}

The computation at the next order is straightforward and we give here only the main lines. Substituting the expansions (\ref{expansion}) into Eqs.~(\ref{model-adim2}), we get at order $\epsilon^2$ the following system of equations
\begin{subequations}
\begin{align}
H_2' &=-\varphi_2, \quad \varphi_2'=\frac{\varphi_1^2}{2}-H_1^2-P_2+H_2 , \\
P_2' &= \frac{3\, \sigma^3}{H_{\infty}}\, H_1^2-\sigma^3\, H_2,
\end{align}
\end{subequations}
where we have used $H_{\infty}\ll 1$. Eliminating $\varphi_2$ and $P_2$ and using Eqs.~(\ref{H1-phi1-P1}), we get
\begin{equation}
H_2''' + H_2'+\sigma^3\, H_2= \frac{3 \sigma^3\, \tilde{c}_1^2}{H_{\infty}} e^{-2\sigma Z}.
\end{equation}
Considering again that $\sigma$ diverges when $\text{Ca}\to 0$, the solution reads
\begin{equation}
H_2 = \tilde{c}_2\, e^{-\sigma Z} - \frac{3\, \tilde{c}_1^2}{7\, H_{\infty}} e^{-2\sigma Z}.
\end{equation}
Therefore, using the first of Eqs.~(\ref{expansion-H-phi}) and Eqs.~(\ref{H1-phi1-P1}), we obtain
\begin{equation}
H= H_{\infty} + \epsilon(\tilde{c}_1 + \epsilon \tilde{c}_2)\, e^{-\sigma Z} - \epsilon^2 \frac{3\, \tilde{c}_1^2}{7\, H_{\infty}} e^{-2\sigma Z}.
\end{equation}
By posing $X \equiv \epsilon\, e^{-\sigma Z}$, this relation can be inverted. At the leading order, we have
\begin{equation}
e^{-\sigma Z} = \frac{H-H_{\infty}}{c_1} + \frac{3\,(H-H_{\infty})^2}{7\, H_{\infty}\, c_1},
\end{equation}
where $c_1 = \epsilon \tilde{c}_1$. Equivalently, we have
\begin{equation}
Z_{\text{d},\text{i}} = Z_0 -\frac{1}{\sigma}\ln(H-H_{\infty})-\frac{1}{\sigma}\ln\left[1+ \frac{3(H-H_{\infty})}{7\, H_{\infty}}\right].
\end{equation}
The matching with Eq.~(\ref{eq-static-inner-bis}) must be done numerically. We find
\begin{align}
\label{barR-hs-z0-3}
\bar{R} &\simeq 1+2.14\, \text{Ca}^{2/3},\quad H^{\star} \simeq 3.36\, \text{Ca}^{2/3}, \nonumber \\
Z_0 &\simeq \bar{c} + 0.62\, \ln(\text{Ca})\, \text{Ca}^{1/3}.
\end{align}
The matching with the outer solution (\ref{eq-dynamic-outer}) is similar to what has been done at order $\epsilon$ so that, far enough from the fiber, the dynamic meniscus behaves as
\begin{equation}
z_{\text{d}}\simeq \mathcal{R} K_0(r/\ell_c), \quad \mathcal{R} = R\left(1+2.14\, \text{Ca}^{2/3}\right).
\end{equation}
The relative difference with the static meniscus is given by $(\mathcal{R}-R)/R \simeq 2.14\, \text{Ca}^{2/3}$. The coefficient is closer to the one obtained numerical compared to the computation at order $\epsilon$ even if it is still too small.

\begin{figure}[t]
\centering
\includegraphics[width=\columnwidth]{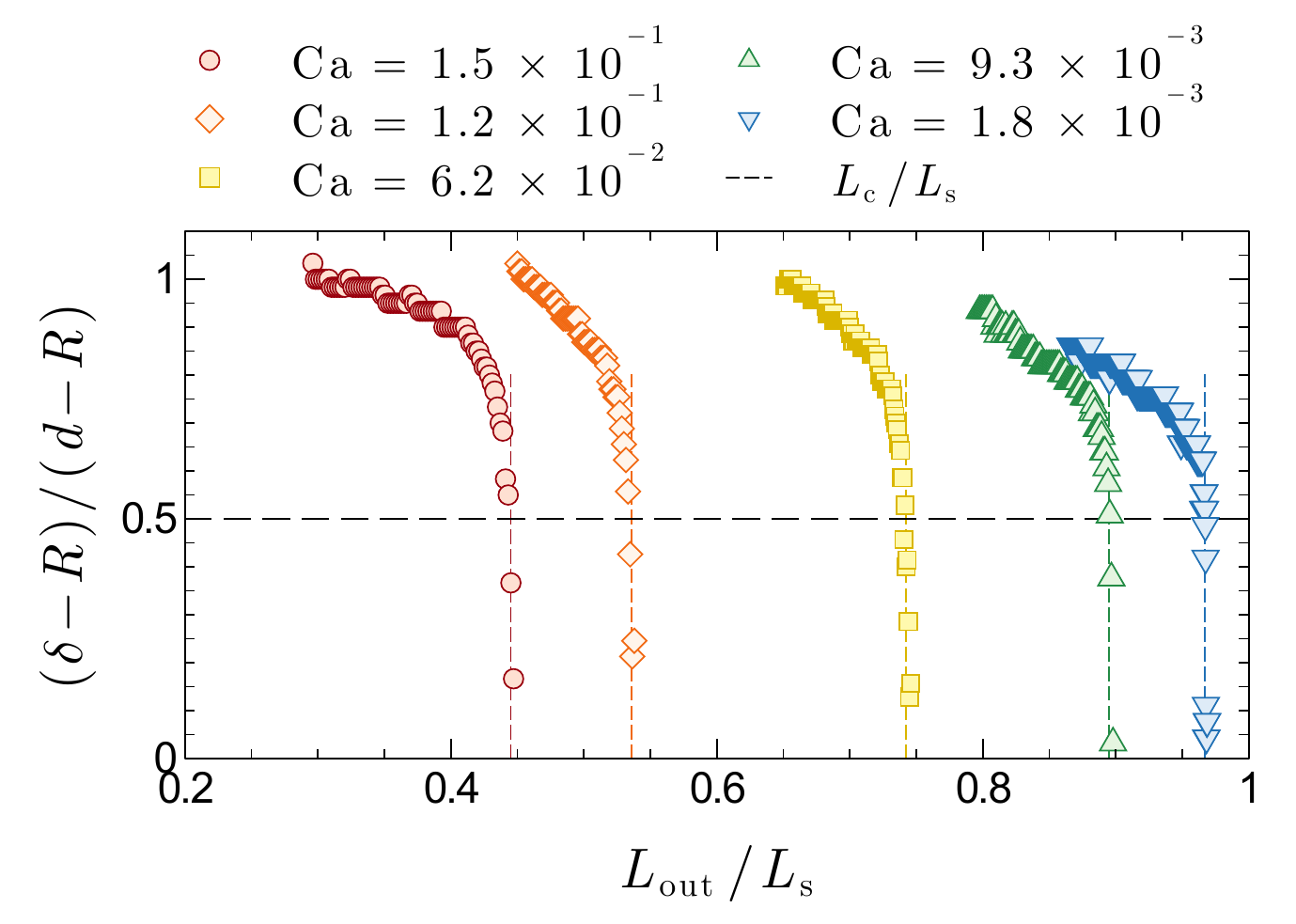}
\caption{Evolution of the rescaled deflection $(\delta-R)/(d-R)$ for various capillary number $\text{Ca}$ as a function of the ratio $L_\text{out}/L_{\text{s}}$, where $L_{\text{s}}=67$ mm is the critical dry length at which the fibers quasi-statically coalesce and $L_\text{out}$ is the vertical distance between the fiber clamped ends and the liquid bath ($R=50$ $\mu$m, $d=0.41$ mm and $\mu= 0.96$ Pa s). The larger $\text{Ca}$, the smaller the coalescence length, indicating an increase in the capillary force at the interface. The transition appears also to be subcritical as in the quasi-static case: the fibers deflect to roughly half their initial distance before snapping to contact.}
\label{figSI:bif}
\end{figure}
 
\section{Linear superposition}

It is commonly assumed that the static meniscus around two fibers is given by the linear superposition of the static meniscus around a single fiber~\cite{cooray2012capillary}. This assumption has been tested in the dynamic, stationary regime by experimentally measuring the profile of the meniscus in the $(x,z)$ plane containing the fibers, in the case of one and two structures withdrawn at the same speed. Figure~\ref{figSI:superpo} and Fig.~3(c) of the main text show that the profile $z_2$ of the dynamical meniscus around two fibers, respectively at $\text{Ca} =0.024$ and $\text{Ca} =0.012$, is given in good approximation by $z_2(x,y)=z_1(x-d,y) + z_1(x+d,y)$ where $z_1$ is the profile of the dynamical meniscus around a single fiber centered at $x=y=0$.

\section{Subcritical bifurcation}

When two fibers, initially separated by a distance $2d$, are removed at a given speed from a liquid bath, the distance $\delta$ between the fibers along the air-liquid interface can be measured as a function of the vertical distance, $L_\text{out}$, between the fiber clamped ends and the liquid bath which increases during an experiment. Representative bifurcation diagrams are shown in Fig.~\ref{figSI:bif} for five values of $\text{Ca}$ with $R/d=0.12$. In these diagrams, the distance $\delta$ is rescaled so that it varies between 1 and 0. When $L_\text{out}$ is large enough and reaches a critical value, $L$, the fibers snap to contact when they are deflected to approximately half their initial distance as in the quasi-static case~\cite{siefert2022capillary}. The critical length, $L$, is given in good approximation by Eq.~(7) of the main text provided $\text{Ca} (R/d)^{3/2}$ is smaller than roughly $3\times 10^{-3}$, i.e. $\text{Ca} \lesssim 7\times 10^{-2}$ in these experiments, see Fig.~4 of the main text. 

\section{Supplemental Movies}

\noindent {\bf Supplemental Movie 1}: Retraction from a silicon oil bath ($\mu = 0.96$ Pa s) of two identical glass fibers of radius $R=50$ $\mu$m separated by a distance $2d=0.82$ mm. For different length $L$ of the fibers, there is a critical capillary number $\text{Ca}$ above which they coalesce. A side-by-side comparison of experiments performed just below the coalescence threshold (left) and above it (right) is presented for 3 different fibers lengths ($L=64.5$, $50.5 $ and $30.3$ mm).

\vspace{2mm}

\noindent {\bf Supplemental Movie 2}: Intricate coalescence mechanism for short fibers ($L=28.4$ mm, $R=50$ $\mu$m, $2d=0.82$ mm and $\mu = 0.96$ Pa s) : below a critical $\text{Ca}^{-}$, they do not coalesce, for intermediate value of $\text{Ca}$ they do coalesce but above another critical $\text{Ca}^{+}$, a liquid column is pulled out between the fibers and a liquid drop is captured, but the fibers do not snap to contact.

\bibliographystyle{apsrev4-2}
\bibliography{reference}

\end{document}